# Meta Analytic Data Integration for Phenotype Prediction: Application to Chronic Fatigue Syndrome

Madhuchhanda Bhattacharjee

School of Mathematics and Statistics, University of Hyderabad, Hyderabad-500046, India

Email address: chhanda.bhatta@gmail.com

Running title: Predictive model using meta analytic data integration

**Summary**

Predictive modeling plays a key role in providing more accurate prognosis and enables us to take a step closer to personalized treatment. We identified two potential sources of human induced biases that can evidently lead to disparate conclusions. We illustrate through a rather complex phenotype that robust results can still be drawn after accounting for such biases.

Often predictive models build based in high dimensional data suffers from the drawback of lack of interpretability. To achieve interpretability in the form of description of the organism level phenomena in term of molecular or cellular level activities, functional and pathway information is often augmented. Functional information can greatly facilitate the interpretation of the results of the predictive model.

However an important aspect of (vertical) data augmentation is routinely ignored, that is there could be several stages of analysis where such information could be meaningfully integrated. There is no know criteria to enable us to assess the effect of such augmentation. A novel aspect of the proposed work is in exploring possibilities of stages of analysis where functional information may be incorporated and in assessing the extent to which the ultimate conclusions would differ depending on level of amalgamation.

In order to boost our confidence on the key biological findings a first level of meta-analysis is carried out by exploring different levels of data augmentation. This is followed by comparison of predictive models across different definitions of the same phenotype developed by different groups, which is also an extended form of meta-analysis.

We have used real life data on a complex phenotype to illustrate the above. The data pertains to Chronic Fatigue Syndrome (CFS) and another novel aspect of the current work is in modeling the underlying continuous symptom measurements for CFS, which is the first for this disease to our knowledge.

**Keywords:** Phenotype prediction, functional data, supervised-PCA, cross validation, meta-analysis, Chronic fatigue syndrome.

## 1. Introduction

In this exercise consider predictive model building for continuous phenotype based on genomic data. Such models often become difficult to interpret in the biological context thus common practice is to augment functional information. However we routinely fail to realize that there could be several alternate steps in the model building effort where such information can be utilized and as can be expected ultimate functional interpretation could and does depend on where such data was incorporated. Our objective would be to explore whether robustness of final findings can be achieved when faced with such alternate choices of integrating same data. Apart from the continuous nature of the phenotype that we target to model and predict, clinical complexity of such phenotypes also mean alternate definition of measurement methods for such phenotypes. Common experience is that prominent biological pathways identified behind such phenotype depend on how the phenotype was defined.

Thus we have identified three issues that have emerged as key challenges in carrying out reproducible research in this field. First and foremost is in realizing that model fit is not enough, in order to bring benefits to larger population more efforts must be put in predictive model building. With sufficient degree of complexity a model might fit data in hand perfectly but might be far from satisfactory when it comes to predicting unseen (i.e. outside the sample) data.

Secondly, with the wealth of availability of data, routinely data is to augment from various sources. As in horizontal data integration one can consider multiple type of useful data, similarly for vertical data integration often there are multiple stages in the analysis pipeline where one might amalgamate such information. For horizontal data integration it is customary to report findings based on different data sources however rarely it is investigated how robust are the final findings in view of alternate methods of vertical data augmentation.

Third aspect we would like to draw attention to is that of phenotype definition which plays highly critical role for complex diseases. Many diseases/syndromes occur in multitude of clinical symptoms leading to multiple ways of defining them. Once again a relevant issue would be to assess the impact of alternate definitions of the phenotype on our conclusions. Here our effort would be in providing a solution that will enable us identify predictors of complex phenotypes that would be robust against both alternate choices of data integration and alternate definitions of phenotype.

With the increasing availability of high dimensional (genomic) data on subjects, it has become essential to device techniques to relate such high-dimensional genetic or genomic data to various (clinical) phenotypes of patients. Modeling and/or analysis of high dimensional data, (e.g. genome-scale genetic data, genomic/transcriptomic data, new-generation sequencing data, etc,) has received great attention for more than a decade. However only recently predictive model building based on high throughput data has gained momentum. Recent focus of research in this area is from the urge to bring benefits of advanced research from bench to bedside and provide personalized deductions from models.

High dimensional data generates a common problem to both model fitting and predictive model building efforts, which is to identify key variables. As the name suggests data from high-throughput techniques are often quite large in size, thus initially substantial effort might have to be put in to derive usable/interpretable variables. Depending on the technology used to capture the covariate information the explanatory data could be binary (genotype data), discrete (e.g. sequence information), continuous (inclusive but not limited to gene expression). Irrespective of the type of experimental data the end model still might not be readily interpretable.

It is possible to derive biologically meaningful covariates based on the original ones that are also smaller in number. External knowledge from databases like Gene Ontology (Ashburner et al., 2000) has been used by Chen (2008), from KEGG (Kanehisa et al., 2008) by Li and Li (2008). Further work on this can be found in Binder & Schumacher (2009) and Pan et al. (2010). Thus a step towards easier interpretability would be to use such functional data. We attempt to augment additional knowledge from existing databases, in particular the Gene Ontology database, to provide better interpretability of such models.

However even if we identify some specific source of data to amalgamate, there might still be multiple stages in the analysis pipeline where such data can be augmented meaningfully. Our experience of decades of data analyses of real data has taught us that the results based on different ways of data integration often leads to different biological results. We investigated effects of such alternate manners of data augmentation and propose a consensus method to arrive at robust conclusions. To the best of our knowledge this has not been attempted before (even for model fitting, let alone for predictive model building).

Predictive modeling has reasonable literature when the response variable is dichotomous/ categorical in nature, where primarily methods for classification are employed/extended. Classification is a relatively easier problem, since as long as the conclusion is on the correct side of cut-off, a decision is considered to be correct and the margin of correctness is deemed irrelevant. Thus for same problem it is not surprising to achieve considerable accuracy in classification, say 70-80%, whereas when prediction of the continuous phenotype is attempted, association measure between predicted and observed quantities might be rather small ranging 0.10-0.20 (on the scale from -1 to +1) (see Bhattacharjee 2013, Bøvelstad & Borgan 2011). Thus for ease of analysis and/or ease of interpretation often the underlying response is recoded into categories. However such categorization of the continuous phenotype leads to loss of information, additionally many situations require dealing with the actual phenotype in its continuous form.

An additional layer of complicacy noticed for complex phenotypes is varied manners/definitions in which they can be described. Often these diseases manifest in a range of symptoms and consequently there could be multiple ways of gathering information and quantification, as we will see in the real life example used for illustration here (see also Haibe--Kains et al Nature, 2013, for an example from pharmacogenomics).

In this work we will explore meta-analytic approaches for vertical data integration which enables robust prediction of phenotypes. The proposed concepts will be illustrated by probabilistic modeling of real life data on complex phenotypes and augmented of high throughput data. We would use a well-known data set on chronic fatigue syndrome (CFS). For 79 individuals binary classifications of individuals according to CFS and otherwise were proposed by Reeves et al. (2005). So far this is the most commonly analyzed phenotype data for CFS based on this data. However it is the (continuous) symptom scores that were measured originally. Our objective is to build predictive models based on genomic data with additional data augmented from the Gene Ontology database.

## 2. Materials and methods

### 2.1 *Data*

The gene expression data and the CFS symptom scores used in here were made available during the CAMDA 2006 and CAMDA 2007 conferences. The data pertains to self-administered questionnaires filled in by 227 subjects from whom blood samples were drawn subsequently for laboratory analysis.

Subject recruitment, clinical evaluation, laboratory tests and their classification were described previously (Reeves et al 2005). As reported therein, 227 subjects were recruited from Wichita, Kansas, USA as part of a two-day in-hospital evaluation of unexplained fatigue. During the two-day hospital stay, symptoms and exclusionary medical and psychiatric conditions were re-evaluated for all 227 subjects. Following the two-day hospital study, all subjects were classified based on all aspects specified in 1994 CFS case definition (Fukuda 1994). The symptom information was recorded according to the Medical Outcomes **S**hort-**F**orm (SF36), **M**ulti-**D**imensional-**F**atigue Inventory (MDF) and **C**enter for **D**isease **C**ontrol (CDC) Symptom Inventory. The symptoms include measures on the functional impairment, fatigue and accompanying symptom complex that characterize CFS.

Following this classification, 124 subjects were excluded because of medical or psychiatric exclusionary conditions or insufficient criteria to classify as CFS. Usable microarray data were available for 79 of the 103 remaining subjects. Based on earlier work of Reeves et al (2005) these 79 individuals can be classified as CFS (39 subjects) and non-fatigued (NF, 40 subjects). The demographic characteristics along with type of disease onset (gradual vs. sudden) of subjects in this study are given in Table 1. We note from this table that the two groups of patients (viz. CFS and NF) appear similar with respect to the distributions of several important relevant factors, like age, gender, BMI etc. Thus this eliminates possibility of confounding factors affecting the conclusions of our analyses.

However since degree of affect of the disease on a subject vary substantially thus it would be an over simplification to dichotomize the disease status,. Even worse there are phenotypes, like CFS, which could be hard to even define and are without any clinical indicators that can confirm the presence of the disease. Therefore unlike previous studies where only the CFS and NF characterization of subjects have been used, in this study we would like to target modeling the underlying disease score.

Further complications arise in regard to diseases like CFS where there are alternate opinions as to how to capture the severity/effect of this disease on a subject. We have used three prominent notions of quantifying this disease, namely, SF36, MDF and CDC symptom inventory mentioned above.

Direct aggregation method is employed to derive the symptom scores for both SF-36 and MDF methods. For CDC Symptom Inventory was used, methods recommended by Wagner et al (2005) was used. In this methods the presence of a symptom along with its frequency and severity were used (multiplicatively) to yield individual symptom level score and these scores were then added to obtain the combined symptom score for each subject. Further details are available in the supplementary materials. The score based on CDC symptom inventory can be further subdivided into the two categories, viz. directly related to CFS and otherwise relevant. Table -2 contains the symptoms covered under each of these methods of defining CFS.

As mentioned above, our objective here would be to model and predict the symptom scores and not the dichotomized disease status (which is more commonly done). Given the complexity of a disease like CFS and the diversity of symptoms in which it manifests, one would expect it would be natural to consider an appropriate measure for this disease than denoting people as having the disease or not. Our literature search revealed only one another work which comes close to this objective (Smith et al. 2009).

Figure 1 presents some descriptive summary of these symptom scores. As can be seen none of three prominent phenotype measures, viz. the Total-score, SF36 and MDF corresponds to clear relationship with the binary phenotype status. The CFS related score however has nearly distinct distributional behavior for the CFS and NF individuals, based on the current data set.

As mentioned in the introduction, interpretation of predictive models is a challenging job. It is even more so for a complex disease like CFS where the symptoms are diverse and overlapping with many other diseases. Thus additional knowledge on the covariates could shed light into the effects and interpretations of the predictors (along with their effect estimates) on to the response variable. There are few well-known data bases that contain information of biological processes, molecular functions of genomic entities, e.g. Gene Ontology, KEGG databases. We would utilize functional information on the genes from the Gene Ontology database (Ashburner et al. 2000), which has larger coverage in terms of number of genes. This information can be utilized at different levels of modeling, where one can start with expression profiles of biological functions instead of individual gene expressions as predictors, also on the other hand one can follow up the influence of the biological profiles at the end of gene expression based predictive modeling.

*2.2 Analysis Methods*

Using gene expression data with various choices of predictive models (e.g. Multiple Regression, Generalized Linear Model, Cox-proportional hazard model, etc) and variable selection techniques (e.g. supervised PCA, LASSO, ridge regression) have been reported in Bhattacharjee (2013). It appears that the choice of variable selection technique seems to have greater effect on the prediction quality than the predictive model itself (similar findings were reported for other data by Bovelsatd et al (2011).

Also let’s remind ourselves that our objective here is for a given phenotype to arrive at a consensus set of (biologically interpretable) predictors taking into account known variabilities like data augmentation choices and phenotype definitions. If we further introduce widely different modeling techniques, which potentially optimize different characteristics, then it is unlikely we would reach such a consensus.

For our data related to CFS the SPCA seemed to perform the better than the other models. Note that we do not restrict to the use of first principal component only. To quote from Paul et al (2008), “Typically, we use just the first or first few supervised principal components”. It appears unfortunately in many applications people pay attention to only half of that recommendation. Most data that we analyzed required many more than “first” PC to achieve reasonable performance. Additionally unlike Chen (2008) instead of fixing number of PCs, a range of number of PCs was explored. For each such choice additionally various thresholds as cutoff for correlation in the supervised stage of the SPCA were used. This resulted in over 1000 models being explored for each choice of data augmentation method and definition of phenotype.

Following SPCA based dimension reduction we implemented multiple regression based predictive modeling. The optimal tuning parameters are to be determined based on a 5-fold cross-validation. Therefore methods used to analyze the data sets in this paper are (i) principal component analysis (PCA), (ii) supervised principal component analysis (SPCA), (iii) linear models and (iv) cross-validation. Further details on each of these have been provided in the supplementary material.

*2.3 Functional data incorporation*

In the context of CFS, attempting to use **G**ene **O**ntology (GO) data (Ashburner et al. 2000) is not a novel idea. Emmert-Streib (2007) used 12 pathways and their expressions to make predictions about the modification of pathways due to pathogenesis. However given the diverse symptom manifestation of CFS, it is a long foregone conclusion that possibility of characterizing

CFS by handful of genes or functions is futile. We are also not attempting to identify or order key pathways or biological functions as in Chen et al. (2008), thus it allows us to use as many functional categories relevant for the disease as necessary.

Thus retaining as much of the relevant ontology information as possible, we propose the first model where using the entire gene expression matrix and corresponding mapping of functions, functional level expressions are estimated. These pathway level expressions are used as predictors in SPCS and subsequent predictive models, denoted as Model-1: "GO-start" (indicating GO information being used from the very start of analysis).

Chen et al. (2008) proposed a method for assessing significant association of gene-ontology defined sets of genes with continuous outcomes. They used the screening/variable selection step of the SPCA method first before utilizing the Gene ontology information. This is because the entire set of genes related to a particular biological function would be affected by variation unrelated to outcome.

Our objective here is to produce predictive model as accurate as possible which would also enable us to carry out interpretations simultaneously. Thus in a spirit similar to that proposed by Chen et al (2008) we consider a model where the Gene ontology information would be used only on the selected set of genes that are screened through the supervised step of the SPCA. Since the GO information is utilized midway in the analysis pipeline we would refer this second model as Model-2: "GO-mid".

Lastly we propose to use the more commonly used approach of using gene –expressions for deriving the predictive model and then following up with interpretation using functional information. In this set up the SPCA and predictive modeling is carried out using the gene level expression values and only at the end stage of analysis functional information is incorporated /

projected on to the genes (or their effects on the phenotypes). We would refer to this model as Model 3: “GO-end”.

Interestingly both in Model-1 and Model-3 gene ontology information on all genes are used (either in building the predictive model or in subsequent interpretation). However in Model-2 since a subset of genes is selected (for each fold and each threshold value) the GO information on these genes is used in the subsequent analysis and interpretation.

In the following we present a flow-diagram depicting the analysis pipeline under different modeling strategy proposed by us (Figure-2).

**3. Results**

In all modeling efforts 5-fold cross-validation was carried out. In each step of analysis we ensured that information from the learning/training set is not leaked into test set. For example the supervised step wherein key variables are selected based on their relevance is carried out based on data on subject from learning set only. Thus often 5-fold cross validation leads to 5 possibly different lists of variables. This error can and is common to occur unintentionally in many analyses (see Bair 2004 for an example).

As mentioned above for each phenotype, each choice of GO data incorporation level and each fold, more than 1000 models were applied. For every model using the predictive values for the test data predictive correlations were obtained. We used these to compare performances of different models. For each of the three types of GO-input method and four types of phenotype, the model with highest correlation with the phenotype was deemed to be the best model.

To identify relative key GO terms the fold level coefficients estimates were consolidated for each GO term using average. This method could be applied for GO-Start and GO-Mid data augmentation techniques. Once again the absolute coefficients for GO-terms were used to measure their effects on a predictive model.

However GO-End produces coefficients at gene level and further mapping between genes and GO-terms were used to derive GO-term level coefficient. This however leads to only marginal estimates for GO-terms and not joint estimates as obtained under GO-Start and GO-mid. Thus in further analyses of identification of GO terms GO-End results were not included due to lack of compatibility.

We observed that the best performing model thus identified may not be consistent across phenotypes and even GO-input method in terms of coefficients of the GO-terms. Thus instead of restricting ourselves to top few models and top few functions, we assessed the consistency of effect estimates for all GO-functions considered here across all 1026 models under each modeling scenario and each of four phenotypes. In most cases although the top performing models may not correlate highly, we could identify models with similar predictive performance and also exhibiting high association with models for other phenotypes/GO-input method. In Table-3 we presented results for the best model and a closely performing alternate model. The predictive correlation measure for model performance and also pair wise association measures bewteen these 16 models are presented therein. Note that results from GO-End strategy have been omitted due to compatibility issues mentioned earlier.

We can further reduce these selected 16 models with one model for each phenotype while still preserving reasonable prediction performance quality. In table-4 we present the performance, pair wise similarity of GO-coefficients and concurrence of GO-terms in top 20% (in absolute magnitude) in the predictive models. This indicates that not only a large set of functionalities appear to be common under these models but their manner of affecting the phenotype is also very similar.

Most of these models are relatively sparse and although we considered the top 20% coefficients based on absolute magnitude (i.e. 156 functions), the effective number of non-zero coefficients could be even less. We could identify 69 GO-based biological functions that consistently showed up across all four phenotype definitions, two data augmentation methods and in terms of their high relevance in predicting the respective phenotype. We present such a list in Table-5.

We will reiterate that objective of this exercise was not necessarily to derive the optimal modeling strategy for predicting this particular instead illustrate a method to account for some known but often ignored sources of variation. However there is still a need to produce some assessment whether the predictive qualities achieved here were of any significance or not.

Unfortunately to the best of our knowledge there is no previous record of modeling this disease in continuous phenotype form. Thus whether the performances of the top models are adequate or not would require some alternate means of investigation. We return our focus to the relatively more studies binary representation of the phenotype. One possibility would be to use cut-off on the estimated phenotype values to conclude about the binary disease status. However we have to recognize that it would be futile to try and implement that strategy for each of the four phenotypes. Thus we used the phenoype-1 which is based on "All-symptoms" from the CDC-symptom inventory, which is the natural choice given both the binary and this particular continues measure were proposed by same group. It should be mentioned here that even when the true knowledge on the continuous phenotypes is available, they do not accurately identify the underlying dichotomized disease status. The accuracy of phenotype-1 based on CDC's "All-symptom" is 93.7% where as that using "CFS-related symptoms" is 98.7%.

However it is also possible to carry out similar modeling strategy can be extended to the binary phenotype also, by implementing SPCA and Generalized Linear Models, where the predictive model is changed to reflect the binary nature of the response function. As before the GO data incorporation can be carried out at various stages.

The results are presented in Table-6 where we notice that even a crude threshold based method (on the continuous prediction) can produce comparable or even better performance than modeling the binary phenotype itself.

The earlier work on such prediction of CFS status based on similar data can found in Bassetti et al.(2006). They report average classification accuracy of 77% on 130 patients using K-means clustering. Given that our focus here has been to estimate the harder problem of continuous phenotype and estimation of binary disease status is only of secondary interest, where the continuous outcome in truth also does not clearly identify the binary classes, we feel it will not be unreasonable to believe that the above performance (in Table-6) is quite encouraging. We need to keep in mind that these are results of out-of-sample prediction and not measures of model fit (of data at hand).

As has been reported in Bhattacharjee (2013) prediction of the continuous phenotype using the gene level expression measurements (and without functional data) reaches correlations about 0.4 using similar modeling strategy. By using threshold method on these predictions an accuracy of 71% can be achieved for the binary phenotype, which is less or comparable to the accuracies achieved here.

Thus if one would wonder if there has been any noticeable deterioration in performance of the predictive model by amalgamating the gene level information to functional level summaries, one would be happy to conclude that neither prediction of the continuous phenotype nor accuracy of prediction of the binary phenotype has been compromised in any way. We should keep in mind that the predictive modeling frameworks have been maintained at comparable level across all these analyses (i.e. use of SPCA and GLM).

**4. Discussion**

Proposed work has several novel aspects, first being those relevant to the specific disease chosen here, namely CFS. Although is CFS is measured using symptom severities, nearly all modeling efforts use dichotomized version of these, thus losing information and also subjective dichotomization leading to lack of inter study comparability. As mentioned before to the best of our knowledge this is the first known effort to model the Chronic Fatigue Syndrome on a continuous scale of measurement. We attempted to model the underlying continuous disease measure for Chronic Fatigue Syndrome, instead of using the dichotomized phenotype. This evidently makes it harder but when compared with approximate accuracy rate of the binary phenotype prediction, the results of continuous phenotype prediction appear encouraging. Additionally the model being predictive provides information on not just the data at hand but also similar unseen data sets.

A more general and novel applicability of our method is that we have identified and rectified against two potential artificial sources of bias often introduced in the analyses by the researchers. The first is in recognizing that functional information can greatly facilitate the interpretation of the results of the (predictive) model. However there is no know criteria to enable us to assess the effect of such augmentation. As we have shown through the lack of consistency among the "Best" models that each such method can very well lead to completely dispersed conclusions in identifying key predictors (as noted by poor or even negative correlations in table-3).

Often it is the researcher and not the biological context that determines which method for data augmentation would be used. We have explored possibilities of stages of analysis where functional information may be incorporated and also assessed whether the ultimate conclusions would differ depending on level of amalgamation. We have shown that with little or no compromise in model performance the difference strategies of incorporating GO data we could identify consistent conclusions, even for an well known complex phenotype like the one considered here.

Accurate phenotype definition is key to reproducibility, as is well understood by now. However complex phenotypes like CFS manifests in a gamete of symptoms and once again it is the researcher who decides and uses specific definition of his/her choice. It is then not surprising the conclusions differ from study to study as we illustrated once again (in Table-3) that even using exactly same modeling framework we might very well make diverse conclusions regarding the key network/biological players. This too can be avoided and common factors identified through consensus method as carried out here may lead to better understanding of the phenotype.

As for extension of the current work, we noticed from figure-1 that it is likely that the underlying phenotype distributions are asymmetric and has longer right tails. A method to model this would be in changing the predictive model from linear model to generalized linear model (e.g. by using a Gamma distribution for response). However although in theory the method should work in practice because of the very nature of this data the standard computational procedure of IRWLS (i.e Iteratively Re-weighted Least Squares) method to obtain the estimates for such a Gamma response model often fail in practice. This it would be worth to investigate in future how to overcome such computational difficulties in modeling with other distributions from the exponential family. A second aspect would be of course to use multiple experimental data, that could help reduce effect of sampling bias in the conclusions. It is rather phenotype specific whether multiple such data sets with stuitable information of genomic and phenotypic data would be available.

*Acknowledgements:*

MB gratefully acknowledges the initial work done by Ms. Anjana Bapat and Ms. Tanvi Vilekar on modeling the continuous scores from this CFS data as a part of their Master's project titled "Survival analysis in patients with Chronic Fatigue Syndrome using clinical and gene expression data" under MB's supervision in the Department of Statistics, University of Pune, Pune, India. She also acknowledges contributions of Prof. S.R. Deshmukh (for participating in an earlier version of the ms) and Dr. Suzanne Hagan (for helpful discussions). The author declares that no sources of financial support were used in preparation of this manuscript.

*Declaration of Interest*

Author declares no conflict of interest.

**FIGURE LEGENDS**

**Figure 1**: Box-plots of the four different types of symptom score according to dichotomized symptom score of NF (non-fatigued) and CFS patients.

**Figure 2**: Schematic diagram of analysis flow chart where gene ontology information is incorporated at different level of analysis yielding different model.

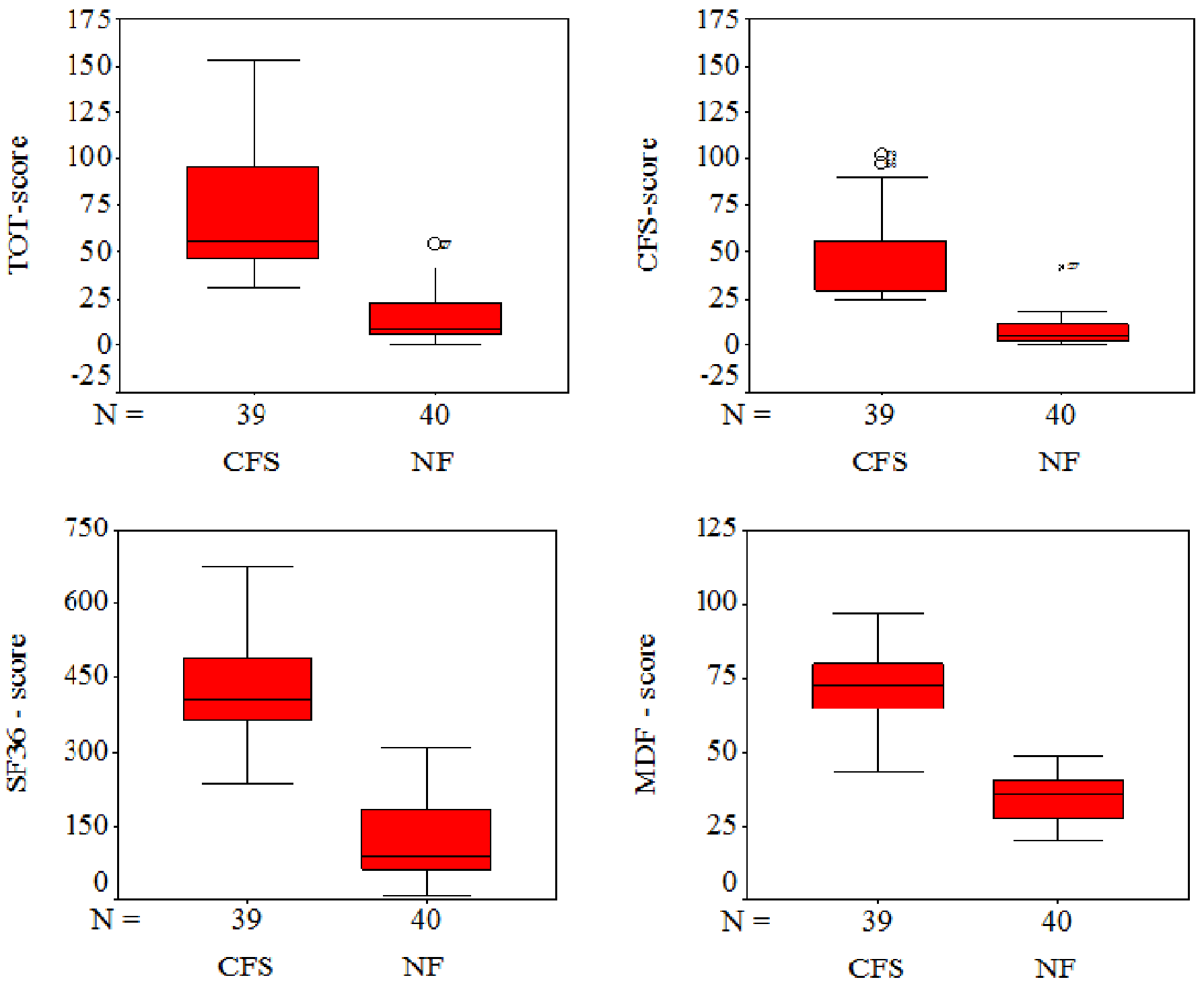

Figure 1

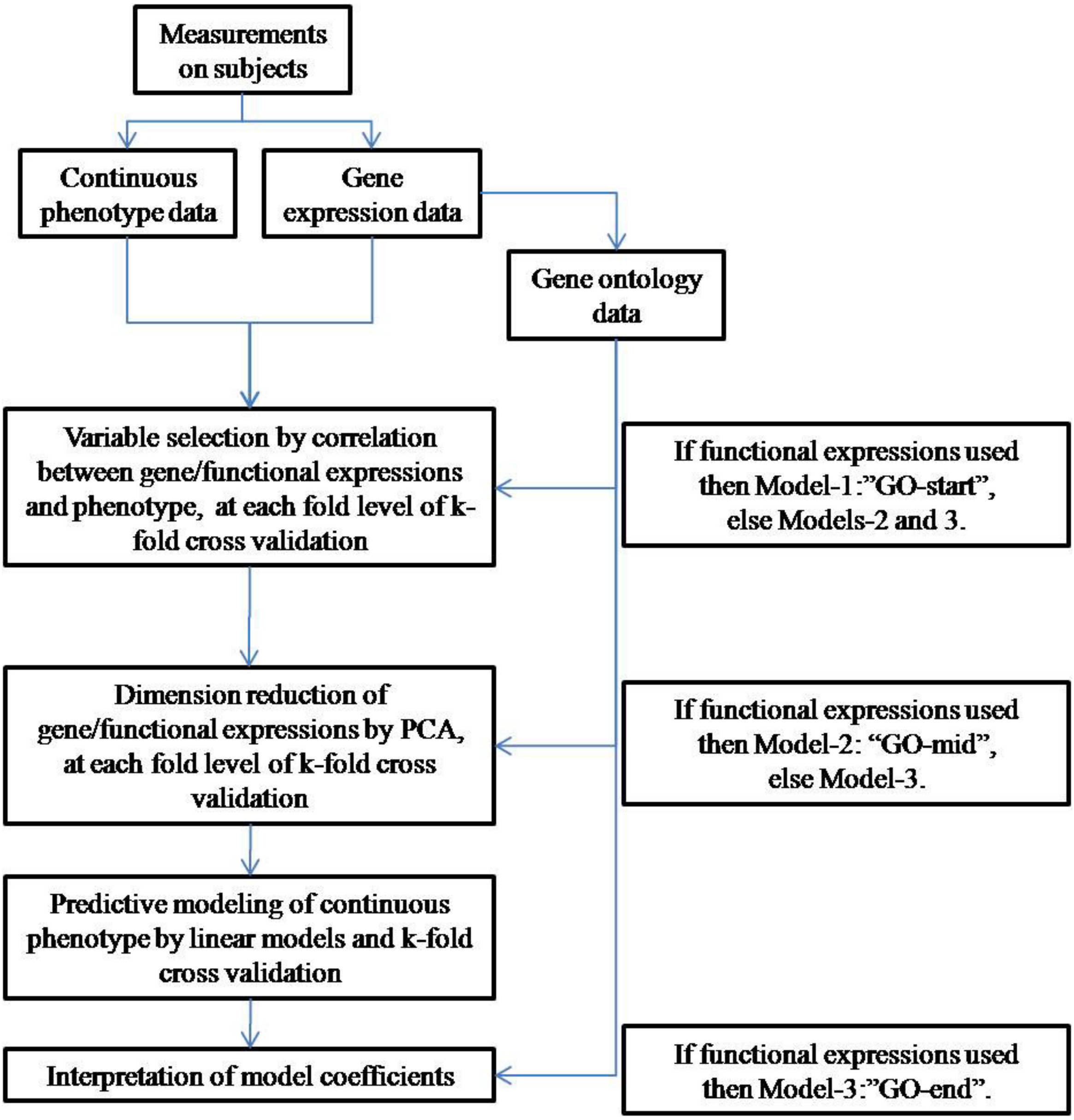

Measurements on subjects
Continuous phenotype data
Gene expression data
Gene ontology data
Variable selection by correlation between gene/functional expressions and phenotype, at each fold level of k-fold cross validation
If functional expressions used then Model-1:"GO-start", else Models-2 and 3.
Dimension reduction of gene/functional expressions by PCA, at each fold level of k-fold cross validation
If functional expressions used then Model-2: "GO-mid", else Model-3.
Predictive modeling of continuous phenotype by linear models and k-fold cross validation
Interpretation of model coefficients
If functional expressions used then Model-3:"GO-end".

Figure 2

**Table 1**: Demographic and other relevant characteristics of the subjects selected for analysis.

| Factor | Categories | NF subjects (n=40) | CFS subjects (n=39) |
|---|---|---|---|
| Age (years) | MQMQM[a] | 31 / 43.8 / 50.5 / 53.3 / 69 | 34 / 47 / 53 / 58.5 / 69 |
| Sex (n) | Female/Male | 31 / 9 | 32 / 7 |
| Race(n) | White/Black/Others | 36 / 2 / 2 | 36 / 1 / 2 |
| BMI | MQMQM[a] | 20 / 26 / 29 / 31.3 / 40 | 23 / 26 / 29 / 31.5 / 40 |
| Onset[b] | Gradual/ Sudden | 13 / 1 | 35 / 3 |

[a]MQMQM represents the minimum, first quartile, median, third quartile and maximum respectively.

[b]Onset represents gradual vs sudden onset of illness. This information is available for all but one CFS subject. For NF subjects, onset information is relevant to only 14 individuals with past report of chronic fatigue.

**Table 2**: Symptoms from CDC inventory to derive the two continuous phenotype measures related to CFS, namely CFS-score and Total-score.

| CDC symptom inventory | | Medical Outcomes Survey Short-Form (SF-36) | Multidimensional Fatigue Inventory (MFI) |
|---|---|---|---|
| CFS related symptoms | Symptoms otherwise relevant to CFS | | |
| 1. Post Exertion Fatigue | 1. Diarrhea | 1. Limitations in physical activities because of health problems | 1. General Fatigue |
| 2. Un-refreshing Sleep | 2. Fever | 2. Limitations in social activities because of physical or emotional problems | 2. Physical Fatigue |
| 3. Memory | 3. Chills | 3. Limitations in usual role activities because of physical health problems | 3. Mental Fatigue |
| 4. Concentration | 4. Sleep Problems | 4. Bodily pain | 4. Reduced Motivation |
| 5. Muscle Pain | 5. Nausea | 5. General mental health | 5. Reduced Activity |
| 6. Joint Pain | 6. Abdominal Pain | 6. Limitations in usual role activities because of emotional problems | |
| 7. Sore Throat | 7. Sinus Nasal | 7. Vitality (energy and fatigue) | |
| 8. Tender Nodes | 8. Shortness of breath | 8. General health perceptions | |
| 9. Headache | 9. Photophobia | | |
| | 10. Depression | | |

**Table 3**: Correlations between estimated coefficients for all 781 GO functions for top performing and alternate models under different modeling methods and phenotypes.

| | Phenotype | GO input level | Model type | 1 | 2 | 3 | 4 | 5 | 6 | 7 | 8 | 9 | 10 | 11 | 12 | 13 | 14 | 15 | 16 |
|---|---|---|---|---|---|---|---|---|---|---|---|---|---|---|---|---|---|---|---|
| 1 | CDC Symptom Inventory, All symptoms | Start | Best | 1.00 | 0.98 | 0.04 | 0.92 | 0.04 | 0.97 | 0.54 | 0.92 | 0.02 | 0.79 | 0.75 | 0.75 | -0.01 | 0.68 | 0.59 | 0.66 |
| 2 | | | Alt. | 0.98 | 1.00 | 0.05 | 0.94 | 0.04 | 0.99 | 0.56 | 0.93 | 0.03 | 0.80 | 0.76 | 0.76 | -0.01 | 0.69 | 0.61 | 0.67 |
| 3 | | Mid | Best | 0.04 | 0.05 | 1.00 | 0.08 | 0.01 | 0.05 | 0.24 | 0.07 | 0.02 | 0.10 | 0.09 | 0.08 | 0.02 | 0.12 | 0.09 | 0.10 |
| 4 | | | Alt. | 0.92 | 0.94 | 0.08 | 1.00 | 0.04 | 0.93 | 0.52 | 0.97 | 0.04 | 0.78 | 0.77 | 0.80 | 0.00 | 0.64 | 0.56 | 0.63 |
| 5 | CDC Symptom Inventory, CFS related | Start | Best | 0.04 | 0.04 | 0.01 | 0.04 | 1.00 | 0.05 | 0.08 | 0.04 | 0.42 | 0.02 | 0.01 | 0.00 | 0.59 | 0.05 | 0.10 | 0.06 |
| 6 | | | Alt. | 0.97 | 0.99 | 0.05 | 0.93 | 0.05 | 1.00 | 0.57 | 0.94 | 0.02 | 0.80 | 0.77 | 0.77 | 0.00 | 0.70 | 0.62 | 0.68 |
| 7 | | Mid | Best | 0.54 | 0.56 | 0.24 | 0.52 | 0.08 | 0.57 | 1.00 | 0.58 | 0.09 | 0.50 | 0.51 | 0.51 | 0.07 | 0.52 | 0.52 | 0.54 |
| 8 | | | Alt. | 0.92 | 0.93 | 0.07 | 0.97 | 0.04 | 0.94 | 0.58 | 1.00 | 0.04 | 0.79 | 0.79 | 0.81 | 0.00 | 0.66 | 0.60 | 0.65 |
| 9 | SF36 | Start | Best | 0.02 | 0.03 | 0.02 | 0.04 | 0.42 | 0.02 | 0.09 | 0.04 | 1.00 | 0.04 | 0.05 | 0.03 | 0.66 | 0.03 | 0.09 | 0.05 |
| 10 | | | Alt. | 0.79 | 0.80 | 0.10 | 0.78 | 0.02 | 0.80 | 0.50 | 0.79 | 0.04 | 1.00 | 0.96 | 0.96 | 0.00 | 0.87 | 0.75 | 0.84 |
| 11 | | Mid | Best | 0.75 | 0.76 | 0.09 | 0.77 | 0.01 | 0.77 | 0.51 | 0.79 | 0.05 | 0.96 | 1.00 | 0.98 | 0.00 | 0.82 | 0.76 | 0.85 |
| 12 | | | Alt. | 0.75 | 0.76 | 0.08 | 0.80 | 0.00 | 0.77 | 0.51 | 0.81 | 0.03 | 0.96 | 0.98 | 1.00 | -0.02 | 0.81 | 0.74 | 0.82 |
| 13 | MFI | Start | Best | -0.01 | -0.01 | 0.02 | 0.00 | 0.59 | 0.00 | 0.07 | 0.00 | 0.66 | 0.00 | 0.00 | -0.02 | 1.00 | 0.00 | 0.04 | 0.02 |
| 14 | | | Alt. | 0.68 | 0.69 | 0.12 | 0.64 | 0.05 | 0.70 | 0.52 | 0.66 | 0.03 | 0.87 | 0.82 | 0.81 | 0.00 | 1.00 | 0.90 | 0.95 |
| 15 | | Mid | Best | 0.59 | 0.61 | 0.09 | 0.56 | 0.10 | 0.62 | 0.52 | 0.60 | 0.09 | 0.75 | 0.76 | 0.74 | 0.04 | 0.90 | 1.00 | 0.95 |
| 16 | | | Alt. | 0.66 | 0.67 | 0.10 | 0.63 | 0.06 | 0.68 | 0.54 | 0.65 | 0.05 | 0.84 | 0.85 | 0.82 | 0.02 | 0.95 | 0.95 | 1.00 |
| Predictive correlations | | | | 0.35 | 0.34 | 0.39 | 0.33 | 0.31 | 0.30 | 0.32 | 0.29 | 0.28 | 0.24 | 0.21 | 0.20 | 0.39 | 0.34 | 0.35 | 0.34 |

a: Performance is measured by correlation between predicted and observed phenotype.
b: Model-type "Best": Model judged to be top based on performance in out-of-data prediction.
c: Model-type 'Alt.": Model close in performance to top models as in (b) however also consistent across GO data incorporation strategy and phenotype definitions.

**Table 4**: Performance measures and similarity measures for the top chosen "Alternate-model", one for each phenotype definition.

| | | CDC symptom inventory | | Medical Outcomes Survey Short-Form (SF-36) | Multidimensional Fatigue Inventory (MFI) |
|---|---|---|---|---|---|
| | | All symptoms | CFS related symptoms | | |
| | | (1) | (2) | (3) | (4) |
| Predictive correlation | | 0.34 | 0.30 | .024 | 0.34 |
| Correlations between the coefficients of the GO-term for the four chosen predictive models. | (1) | 1.00 | 0.99 | 0.80 | 0.69 |
| | (2) | | 1.00 | 0.80 | 0.70 |
| | (3) | | | 1.00 | 0.87 |
| | (4) | | | | 1.00 |
| Number of common GO-terms with coefficients in top 20% for each chosen model[ab]. | (1) | 156 | 147 | 87 | 77 |
| | (2) | | 156 | 89 | 79 |
| | (3) | | | 155 | 113 |
| | (4) | | | | 156 |

a: A maximum of 156 terms, i.e. 20% of the 781 terms considered, can occur, however a particular model may not use all GO-terms.
b: 69 of these terms are common between all four models (see table-5).

**Table 5**: Top pathways common among the chosen four “Alternate models” (from table-3), one for each phenotype.

| GO code | GO Term |
|---|---|
| GO:0007268 | synaptic transmission |
| GO:0055080 | cation homeostasis |
| GO:0030003 | cellular cation homeostasis |
| GO:0055066 | di, tri-valent inorganic cation homeostasis |
| GO:0030005 | cellular di, tri-valent inorganic cation homeostasis |
| GO:0055065 | metal ion homeostasis |
| GO:0006875 | cellular metal ion homeostasis |
| GO:0006874 | cellular calcium ion homeostasis |
| GO:0014070 | response to organic cyclic substance |
| GO:0043269 | regulation of ion transport |
| GO:0030534 | adult behavior |
| GO:0003001 | generation of a signal involved in cell-cell signaling |
| GO:0040029 | regulation of gene expression, epigenetic |
| GO:0007612 | learning |
| GO:0051271 | negative regulation of cell motion |
| GO:0051924 | regulation of calcium ion transport |
| GO:0030336 | negative regulation of cell migration |
| GO:0050905 | neuromuscular process |
| GO:0008344 | adult locomotory behavior |
| GO:0050769 | positive regulation of neurogenesis |
| GO:0006898 | receptor-mediated endocytosis |
| GO:0034101 | erythrocyte homeostasis |
| GO:0031346 | positive regulation of cell projection organization |
| GO:0008277 | regulation of G-protein coupled receptor protein signaling pathway |
| GO:0030111 | regulation of Wnt receptor signaling pathway |
| GO:0030218 | erythrocyte differentiation |
| GO:0046324 | regulation of glucose import |
| GO:0010827 | regulation of glucose transport |
| GO:0018108 | peptidyl-tyrosine phosphorylation |
| GO:0000060 | protein import into nucleus, translocation |
| GO:0043270 | positive regulation of ion transport |
| GO:0048168 | regulation of neuronal synaptic plasticity |
| GO:0050806 | positive regulation of synaptic transmission |
| GO:0045216 | cell-cell junction organization |
| GO:0050885 | neuromuscular process controlling balance |
| GO:0001936 | regulation of endothelial cell proliferation |
| GO:0003073 | regulation of systemic arterial blood pressure |
| GO:0051928 | positive regulation of calcium ion transport |
| GO:0021987 | cerebral cortex development |
| GO:0051952 | regulation of amine transport |
| GO:0006476 | protein amino acid deacetylation |
| GO:0032312 | regulation of ARF GTPase activity |
| GO:0001974 | blood vessel remodeling |
| GO:0002712 | regulation of B cell mediated immunity |
| GO:0002889 | regulation of immunoglobulin mediated immune response |

| GO code | GO Term |
|---|---|
| GO:0007628 | adult walking behavior |
| GO:0045814 | negative regulation of gene expression, epigenetic |
| GO:0016575 | histone deacetylation |
| GO:0006342 | chromatin silencing |
| GO:0032890 | regulation of organic acid transport |
| GO:0001975 | response to amphetamine |
| GO:0014073 | response to tropane |
| GO:0042220 | response to cocaine |
| GO:0033619 | membrane protein proteolysis |
| GO:0007622 | rhythmic behavior |
| GO:0035162 | embryonic hemopoiesis |
| GO:0042274 | ribosomal small subunit biogenesis |
| GO:0050853 | B cell receptor signaling pathway |
| GO:0045730 | respiratory burst |
| GO:0050974 | detection of mechanical stimulus involved in sensory perception |
| GO:0001963 | synaptic transmission, dopaminergic |
| GO:0034605 | cellular response to heat |
| GO:0045686 | negative regulation of glial cell differentiation |
| GO:0014014 | negative regulation of gliogenesis |
| GO:0046325 | negative regulation of glucose import |
| GO:0050910 | detection of mechanical stimulus involved in sensory perception of sound |
| GO:0018198 | peptidyl-cysteine modification |
| GO:0001993 | regulation of systemic arterial blood pressure by norepinephrine-epinephrine |
| GO:0044070 | regulation of anion transport |

**Table 6**: Prediction of the binary phenotype under different schemes of functional data augmentation. The variable selection is done by SPCA and generalized linear model (in particular logistic regression) is used for predictive modeling.

| Method of estimation | Prediction accuracy for GO-input level | | |
|---|---|---|---|
| | Start | Middle | End |
| Cutoff on estimated phenotype Tot-score | 67% | 73% | 71% |
| Logistic regression as predictive model, with correlation used for variable selection | 70% | 63% | 72% |
| Logistic regression as predictive model, with t-test used for variable selection | 72% | 65% | 72% |

**Supplementary Materials :**
**Meta Analytic Data Integration for Phenotype Prediction:**
**Application to Chronic Fatigue Syndrome**

Madhuchhanda Bhattacharjee

School of Mathematics and Statistics, University of Hyderabad, Hyderabad-500046, India

**1. CFS severity score based on CDC symptom inventory**: As reported in Wagner et al (2005) the CDC symptom Inventory collected information on 19 fatigue and illness-related symptoms (Table-2) during the month preceding the interview. By comparing with the symptoms reported in the same paper, we identified 9 CFS-defining symptoms (pertaining to the 8 symptoms mentioned therein, viz. post-exertional fatigue, un-refreshing sleep, problems remembering or concentrating, muscle aches and pains, joint pain, sore throat, tender lymph nodes and swollen glands, and headaches). The CDC inventory also includes diarrhea, fever, chills, sleeping problems, nausea, stomach or abdominal pain, sinus or nasal problems, shortness of breath, sensitivity to light, and depression. We followed the transformation of frequency and severity as suggested in literature. Let,

$N$: Number of subjects in the study

$X_i$: the score for $i$-th subject, $i = 1, \ldots, N$

$M$: the number of symptoms in a particular data collection method, 8 for SF36, 5 for MDF and 19 for CDC inventory (with details presented in table-2).

$S_{ij}$: severity of $j$-th symptom of the $i$-th subject and was measured on a three-point scale ($1$ = mild, $2$ = moderate, $3$ = severe), $i=1, \ldots, N, j= 1, \ldots, M$, $f_{ij}$: frequency of the $j$-th symptom of the $i$-th subject and was rated on a four-point scale ($1$ = a little of the time, $2$ = some of the time, $3$ = most of time, $4$ = all of the time), $i=1, \ldots, N, j= 1, \ldots, M$.

Then for the $i$-th individual the symptom score is obtained using $X_i = \sum_{j=1}^{M} S_{ij} f_{ij}$. As reported in Reeves et al (2005) the intensity scores were transformed into equidistant scores before multiplication (i.e., 0 = symptom not reported 1 = mild, 2.5 = moderate, 4 = severe).

We propose to use the total symptom scores consisting of all the symptoms mentioned in Table-2 above. Additionally, based on the 9 symptoms that are more focused to CFS, another symptom scores will also be modeled. We will denote the overall score as "Tot-score" and the score based on subset of symptoms as 'CFS-score". In addition to these the SF36 and MDF scores (mentioned above) would also be used occasionally.

**2. Principal component analysis:** PCA is a method of transforming the original variables into new uncorrelated variables which are weighted averages of the original variables. Suppose $X_1, X_2, \ldots, X_p$ are $p$ variables of interest. The first principal component $Y_1$, is defined as a linear combination

$$Y_1 = a_1' X = a_{11} X_1 + a_{12} X_2 + \ldots + a_{1p} X_p$$

where, $X = (X_1, X_2, \ldots, X_p)'$ and $a_1' = (a_{11}, a_{12}, \ldots, a_{1p})$. Vector $a_1$ is chosen so as to maximize the variance of $Y_1$ subject to the constraint $a_1' a_1 = 1$. The second principal component $Y_2$ is a linear combination

$$Y_2 = a_2' X = a_{21}X_1 + a_{22}X_2 + \ldots + a_{2p}X_p,$$

which has the greatest variance subject to two conditions: $a_2' a_2 = 1$ and $a_2' a_1 = 0$.

The second condition ensures that $Y_1$ and $Y_2$ are uncorrelated. Continuing on similar lines, the $j$-th principal component is defined. Thus, in finding principal components we are faced with the problem of maximizing a function of several variables subject to one or more constraints. Using the method of Lagrange's multipliers it is proved that $Var(Y_1)$ is maximized subject to the condition when $a_1$ is the eigen vector of S corresponding to the largest eigen value of $S$, where $S$ is the dispersion matrix of vector $X$. Thus, in general, $Y_j = e'_jX$ is the $j$-th principal component, where $e_j$ is the eigen vector corresponding to the $j$-th largest eigen value, $j = 1, 2, \ldots, p$.

In this paper, vector $X$ is the vector of expression values of $p$ genes. The data on gene expression values for $N$ patients are available. If $M$ denotes the $N \times p$ matrix of gene expression values then principal component can be obtained using the singular value decomposition of matrix $M$ (Jolliffe, 2002).

**3. Supervised principal component analysis**: SPCA is PCA applied on the select set of genes, selected on the basis of association of the genes with the outcome variable (Paul et al. 2008). For example, the genes which are highly correlated with the outcome constitute the select set of genes, secondly the genes may be grouped using gene ontology information, or genes in the same pathway, or genes involved in the same cellular process. Since the set of genes is formed using outcome information the procedure is known as supervised procedure.

**4. Linear model:** In a linear model a variable $Y$, which is a response to $k$ factors $U_1, U_2, \ldots, U_k$, is modeled as $Y = \beta_0 + \beta_1 U_1 + \ldots + \beta_k U_k + \varepsilon$. The parameters $\beta_0, \beta_1, \ldots, \beta_k$ are known as regression coefficients and $\varepsilon$ is the random error, assumed to be normally distributed. The regression coefficients are determined so that the residual sum of squares is minimum. The aim of the analysis is to find out the factors which affect response variable significantly. In this paper response variable is disease score (total or CFS), thus $Y_j$ denotes the response of the $j$-th patient. It may be survival time or any continuous phenotype score, $j = 1, 2, \ldots, N$. The factors may be the gene expression values or the first few PC's of gene expression values or these may be PC's corresponding to genes in the select group, selected according certain criteria. We have adopted the latter approach in this paper.

**5. Cross validation:** Cross validation is a method to evaluate the model, by segmenting the data into training and test sets. In *k*-fold validation the data are segmented into *k* sets. Subsequently, *k* iterations of training and validation are performed such that within each iteration a different fold of the data is held out for validation while the remaining *k-1* are used for training or building the model.